\documentclass[journal,onecolumn]{IEEEtran}
\IEEEoverridecommandlockouts
\usepackage{float}
\usepackage[table]{xcolor}
\usepackage{cite}
\usepackage{subcaption}
\usepackage{multirow}
\usepackage{graphicx}
\usepackage{amsmath,amssymb,amsfonts}
\usepackage{algorithmic}
\usepackage{graphicx}
\usepackage{textcomp}
\usepackage{xcolor}
\usepackage{multirow}
\usepackage{tabularx}
\usepackage{url}

\def\BibTeX{{\rm B\kern-.05em{\sc i\kern-.025em b}\kern-.08em
    T\kern-.1667em\lower.7ex\hbox{E}\kern-.125emX}}

    \newcommand*{\affaddr}[1]{#1} 
\newcommand*{\affmark}[1][*]{\textsuperscript{#1}}
\newcommand*{\email}[1]{\textit{#1}}
    
\begin{document}


\title{Predicting Emotions Perceived from Sounds\footnote{This paper is the pre-print of a paper to appear in the proceedings of the IEEE International Conference on BigData 2020 (Workshops) entitled: ``{\it Predicting emotions Perceived from Sounds}.}}


\author{%
Faranak Abri\affmark[1], Luis Felipe Gutiérrez\affmark[1], Akbar Siami Namin\affmark[1], David R. W. Sears\affmark[2], and Keith S. Jones\affmark[3]\\
\affaddr{\affmark[1]Computer Science Department,}
\affaddr{\affmark[2]College of Visual and Performing Arts,} 
\affaddr{\affmark[3]Department of Psychological Sciences}\\
\affaddr{\affmark[1,2,3]Texas Tech University}\\
\email{\{faranak.abri, Luis.Gutierrez-Espinoza, akbar.namin, david.sears, keith.s.jones\}@ttu.edu}\\
}

\IEEEoverridecommandlockouts
\IEEEpubid{\makebox[\columnwidth]{978-1-7281-6251-5/20/\$31.00~\copyright2020 IEEE \hfill} \hspace{\columnsep}\makebox[\columnwidth]{ }}

\maketitle

\begin{abstract}
Sonification is the science of communication of data and events to users through sounds. Auditory icons, earcons, and speech are the common auditory display schemes utilized in sonification, or more specifically in the use of audio to convey information. Once the captured data are perceived, their meanings, and more importantly, intentions can be interpreted more easily and thus can be employed as a complement to visualization techniques. Through auditory perception it is possible to convey information related to temporal, spatial, or some other context-oriented information. An important research question is whether the emotions perceived from these auditory icons or earcons are predictable in order to build an automated sonification platform. This paper conducts an experiment through which several mainstream and conventional machine learning algorithms are developed to study the prediction of emotions perceived from sounds. To do so, the key features of sounds are captured and then are modeled using machine learning algorithms using feature reduction techniques. We observe that it is possible to predict perceived emotions with high accuracy. In particular, the regression based on Random Forest demonstrated its superiority compared to other machine learning algorithms. 

\end{abstract}

\vspace{-0.1in}
\section{Introduction}
\label{sec:intro}
{\it Affective Computing} is a multidisciplinary field including computer science, cognitive science, and psychology~\cite{Tao2005}. From the computer science perspective, it can be considered as a subfield of artificial intelligence also called ``{\it artificial emotional intelligence}'' that focuses on natural interactions between humans and machines. It aids development of tools to recognize affective states and express emotions~\cite{Picard1997}.

Affect representation can be modeled in a 2D space of 1) {\it arousal} (A), which is the level of eventfulness from bored to excited, and 2) {\it valence} (V), which is the level of pleasantness from sad to happy representing the AV space as proposed by Russell~\cite{Russell1980}. There is also a third dimension called ``dominance'' which is the level of control from weak to empowered~\cite{Russell1980}. The dominance dimension in excluded from this work. 
The emotions can be acquired by self-assessment questionnaires using Self-Assessment Manikins (SAM)~\cite{Bradley1994} or physiological signals such as heart rate, skin temperature (SKT) or brain signals using Brain Computer Interfaces (BCIs) such as  Electroencephalogram (EEG).

{\it Emotion recognition} is a task in affective computing, which studies the techniques for identifying emotions from stimuli such as text, picture, audio and video. These artifacts along with their annotations such as emotion and semantics are usually stored in affective datasets. {\it Audio Emotion Recognition} (AER) is a subfield of emotion recognition and includes emotion recognition from music, speech/voice and sample sound/sound event. {\it Sound emotion recognition} is a relatively new field of research and has broad applications from automatic sound design systems to designing robots, as affective companions. Two types of emotions can be considered when someone is listening to a soundscape: ``{\it Perceived Emotion},'' the emotion expressed by the sound source, and ``{\it Induced Emotion},'' the emotion invoked in the listener.

Audio emotion recognition helps to understand the characteristics of audio samples (e.g. music and soundscapes) regarding the emotions which are induced or perceived and therefore, is useful for designing automated sonification frameworks. Sonification is a field of research that aims to convey information using sound which has been applied in Cyber Physical Systems (CPS) ~\cite{2020Iber, 2019Lenzi} or it can be applied in disaster management~\cite{Nguyen2019}. Auditory representation of emotion, which is also called ``sonification of emotion", is  done by two methods of mapping sounds to an emotion space (e.g. AV space)~\cite{2014winters}: Ecological design, which uses acoustic features suggested by the psychological study of musical emotion such as~\cite{Weninger2014}; and computational design, which utilizes automatic feature extraction methods, such as the MIRToolbox~\cite{MIRToolbox}.
Considering emotion recognition, sonification and Cyber Physical Systems, a novel area of research called ``{\it sonification of emotion in CPSs and IoTs}'' can be explored. 

In designing such sonification-based systems for emotion recognition in CPS and IoT platforms, it is important to represent operational events through meaningful sounds that reflect the emotions perceived or induced by the events. The automatic and effective selection of proper sounds in order to represents events in CPSs depends on whether there are {\it common psychoacoustic features} of sounds that could reliably express emotions represented in a dimensional model of affect. 
More specifically, the objective is to explore the possibility of utilizing sound features in predicting induced or perceived emotions.

This paper intends to investigate the prediction of emotions in AV space perceived from soundscapes with machine learning techniques and reports the best features that can be used in sonification of emotion with application in computer science and more importantly CPSs. The key contributions of this paper are as follows:

\begin{itemize}
    \item[--] Compare the performance of several machine learning algorithms to the emotion recognition problem,
    \item[--] Perform feature reductions and capture the accuracy of the models based on the selected features, and
    \item[--] Identify common features deemed to be important for modeling both arousal and valence (AV space).
\end{itemize}

The remainder of this paper is organized as follows: the related work is reviewed in Section  \ref{sec:related}. Section \ref{sec:dataset} briefly represents the dataset used for this experiment. In Section \ref{sec:models}, machine learning techniques studied in this work are briefly explained. In Section \ref{sec:results}, results obtained by models with and without feature reduction/selection are presented. Section \ref{sec:conclusion} concludes the paper and highlights the future  directions. 

\vspace{-0.05in}
\section{Related Work}
\label{sec:related}
%



Fan et al.~\cite{EmoSoundscape} created the Emo-soundscape dataset of audio samples and their perceived emotion to provide a benchmark for Soundscape Emotion Recognition (SER). They evaluated the dataset using Support Vector Regression (SVR) models. For the feature extraction phase, both YAAFE~\cite{Mathieu2010} and MIRToolbox~\cite{MIRToolbox} tools were utilized. Using the mean and standard deviation of the features, they extracted 122 audio features. Finally, they removed the features whose variances were lower than $0.02$ and thus obtained 39 features. They used this 39-D vector of features to train the SVR models to predict arousal and valence and assess them through $MSE$ and $R^2$ as performance metrics. 
They reported $MSE = 0.048$ and $R^2 = 0.855$ for arousal and $MSE = 0.124$ and $R^2 = 0.629$ for valence, respectively. 

Improving their work, Fan et al.~\cite{Fan2018} used deep learning techniques to predict arousal and valence independently. They used the Emo-Soundscapes dataset for their framework. Given that deep learning methods need great amounts of data, they augmented the dataset using a windowing method to increase the number of samples. By using 30 consecutive windows for each augmented sound sample, they ended up with $8,491$ samples, each being $1.393$ seconds long. They used two different techniques for feature extraction. The the first set of features obtained from a deep CNN model applied in audio classification~\cite{Hershey2017}. For generating the second set of features, they utilized YAAFE~\cite{Mathieu2010} and MIRToolbox~\cite{MIRToolbox} and extracted 54 features including: loudness, energy, perceptual spread, perceptual sharpness, spectral flatness, spectral rolloff, spectral flux, spectral slop, spectral variation, spectral shape, temporal shape, zero cross rate, and 13 MFCCs. They extracted features for each window and because there were 30 windows for each augmented sound sample, they had 54x30 features for each augmented sample. 
Furthermore, they used five different models including: 1) CNN trained through supervised fine-tuning, 
2) CNN trained from scratch that included two convolutional layers followed by one dense layer and 54x30 input features, 3) LSTM-RNN trained from scratch with two stacked LSTM units and 54x30 input features, 4) standard SVR, which uses the first set of features as the input, and 5) Radial Basis Function (RBF) kernel and a combination of CNN and SVR, where VGG-like CNN is used as the feature extractor, and its output was fed into the SVR with RBF kernel. 
They achieved the best arousal prediction with $MSE = 0.035$ and  $R^2 = 0.892$ using the CNN trained from scratch and the best valence prediction with  $MSE = 0.078$ and  $R^2 = 0.759$ using fined-tuned CNN. 

Ntalampiras~\cite{Ntalampiras2020} provided a comparison between emotion prediction from singleton soundscapes and mixed soundscapes using a CNN model. The author used Emo-Soundscape dataset and extracted the features from sound samples using log-Mel spectrum~\cite{Kinnunen2004} which is a spectrogram that the frequencies are converted to the Mel scale. He showed that if the feature vectors of mixed soundscapes and its components (singleton soundscapes) are reduced to their three principal components, most mixed soundscapes are located between its original soundscapes. Ntalampiras conducted the experiment using 2D-CNN models first on a subset of the dataset only containing singleton soundscapes, and then on the whole dataset with mixed soundscapes. Ntalampiras reported the best prediction on the dataset containing both original and mixed samples with $MSE$ scores of $0.010$ and $0.016$ for arousal and valence.
\vspace{-0.1in}
\section{Dataset Description and Analysis}
\label{sec:dataset}
In order to conduct our experiment, we used the Emo-Soundscape dataset~\cite{EmoSoundscape}, which consists of two subsets. The first subset contains 600 audio samples categorized into 6 families, 100 samples each based on Schafer's soundscape taxonomy~\cite{Schafer1977} including: natural sounds, human sounds, sounds and society, mechanical sounds, quiet and silence, and sounds as indicators. The second subset contains 613 samples, each being a combination of soundscapes from two or three classes out of the first subset. All these soundscapes are annotated with their perceived emotion. We used the first subset of this dataset for our experiment. As to the features, we extracted 68 features using MATLAB MIRToolbox~\cite{MIRToolbox}.

The MIRToolbox extracts the (psycho)acoustic and musically related features from databases of audio files for statistical analysis~\cite{MIRToolbox2007}. Following Lange and Frieler~\cite{Lange2018}, a total of 68 features were extracted from each sound sample that represent either the arithmetic mean or the sample standard deviation of the frame-based features computed over default window sizes (typically 50 ms for low-level, and 2-3 seconds for medium level features) and a 50\% overlap. These features can be classified according to their (psycho)acoustic family ({\it dynamics}, {\it rhythm}, {\it timbre/spectrum}, {\it pitch}, and {\it tonality}).

\begin{figure}[h]
  \centering
  \includegraphics[width=\linewidth]{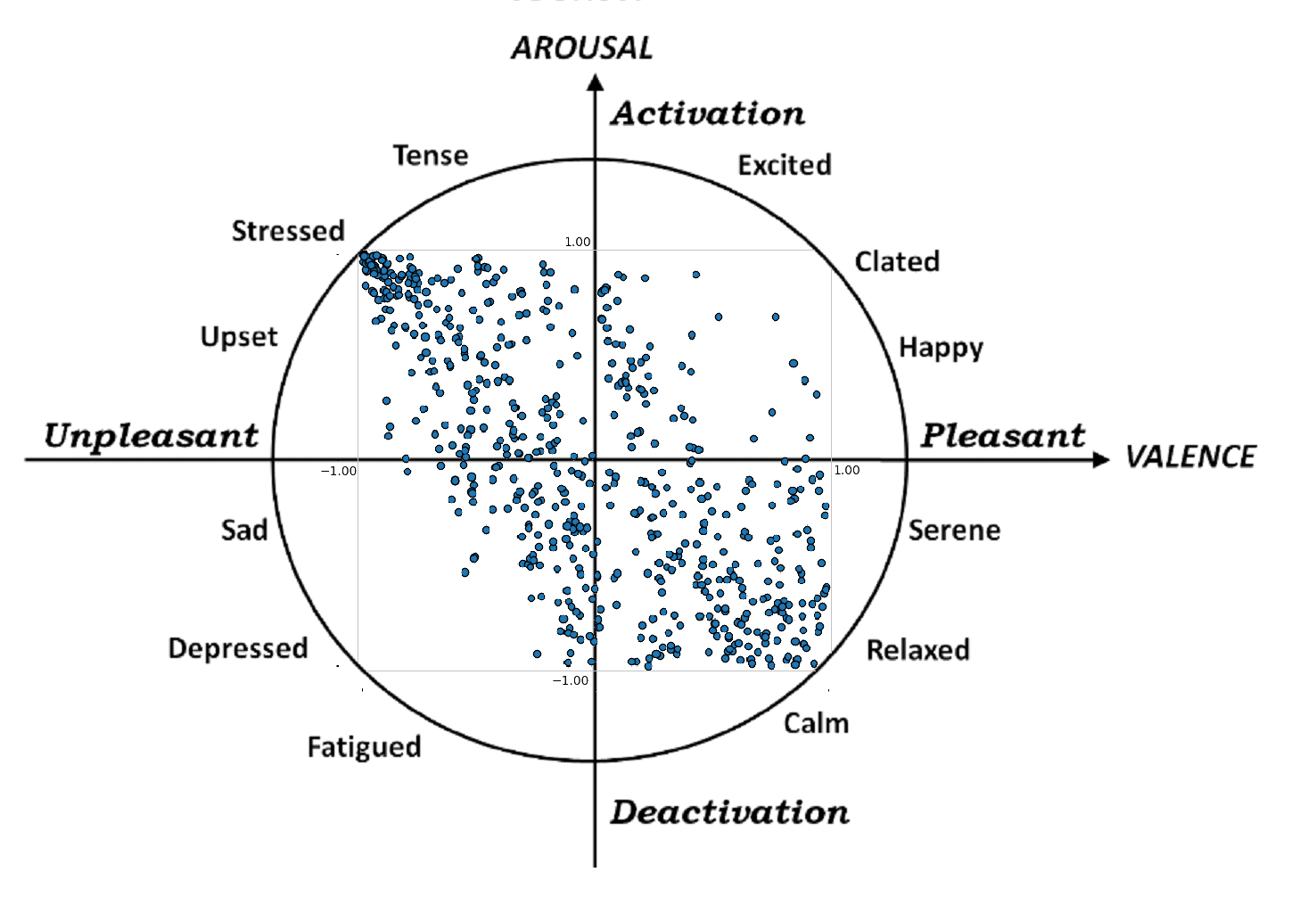}
  \caption{Scatter plot of Emo-Soundscape normalized data points on AV space. The circumplex model is adapted from~\cite{Valenza2012}.}
  \label{fig:valence-vs-arousal}
  \vspace{-0.2in}
\end{figure}

Figure \ref{fig:valence-vs-arousal} shows the scatter plot of normalized values of arousal versus valence for the Emo-Soundscape dataset. There exists a visible trend that suggests a negative correlation between values for arousal and valence. This trend is confirmed after calculating the Pearson correlation $r$ between the two series of values, resulting $r=-0.711$ (p-value $< 0.01$). This correlation suggests that in this dataset, sound stimuli that express excitement are likely to be perceived as unpleasant.

\begin{figure*}[!h]
  \centering
  \includegraphics[width=0.80\linewidth]{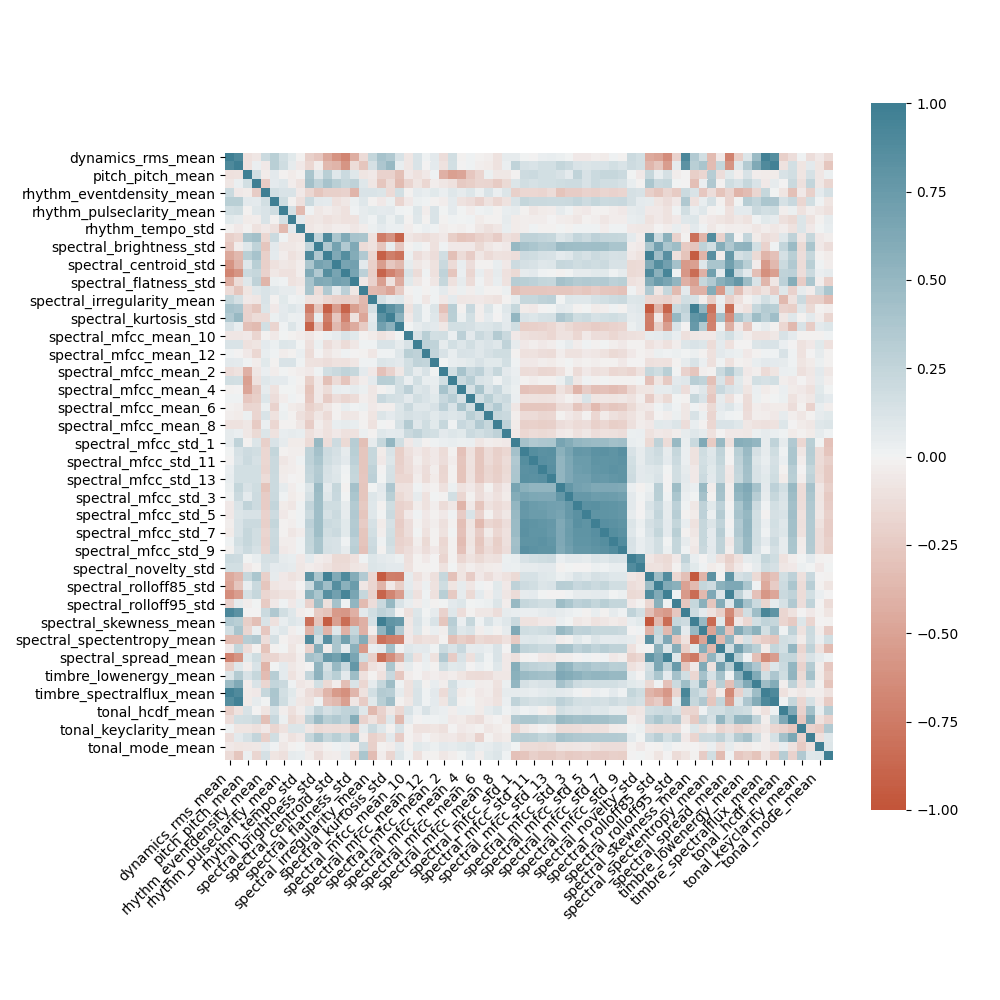}
  \caption{Correlation heatmap between features.}
  \label{fig:corr-heatmap}
  \vspace{-0.2in}
\end{figure*}

Figure \ref{fig:corr-heatmap} shows the heatmap of all pairwise correlations for the 68 features. The heatmap shows that there are more features with positive correlations than negative ones. Also, there exists a very high positive correlation between features dealing with MFCC spectral measures.

\vspace{-0.05in}
\section{Prediction Models}
\label{sec:models}
We used different regression models to predict arousal and valence separately, including four linear models:
\begin{itemize}
    \item[--] $LinearRegression()$, an ordinary least squares linear regression model,
    \item[--] $Lasso()$, a linear model with L1 regularization,
    \item[--] $ElasticNet()$, a linear model with L1 and L2 regularization, and
   \item[--] $SVR(kernel='linear')$, an Epsilon-Support Vector regression model with linear kernel.
\end{itemize}
In addition to these linear models, we explored four non-linear regression models, bringing the total number of prediction models to eight:
\begin{itemize}
    \item[--] $2-layer\ MLP$, a shallow Neural Network model, 
    \item[--] $SVR(kernel='rbf')$, an Epsilon-Support Vector regression model with RBF kernel,
    \item[--] $SVR(kernel='poly')$, an Epsilon-Support Vector regression model with polynomial kernel, and 
    \item[--] $RandomForestRegressor()$, an ensemble estimator that fits a number of decision trees on different subsets of samples. 
\end{itemize}

It should be noted that although deep models such as CNN succeed in achieving remarkable outcomes in the literature (e.g., sequence modeling \cite{DBLP:conf/compsac/Tavakoli20}), we did not examine deep models in our experiment for two reasons. Firstly, the size of the Emo-Soundscape dataset is not large enough (i.e., only 600 samples) for such models and deep models need a fairly large amount of data to perform well. Secondly our feature vector is a 1D vector containing 68 features, which is not suitable to feed into a 2D CNN model. In addition, augmenting data using windowing technique violates the assumption of independence between samples since each of the component sounds from a given sound sample will receive the same rating. Therefore, we decided to examine the dataset without any data augmentation and without any changes in the feature vector using the models explained above.
 
\subsection{Feature Reduction/Selection}
Feature reduction and feature selection help to reduce the number of features and thus the dimensionality of data. Feature selection aims to select a subset of features that represents the entire set of features and helps us to interpret the model. However, feature reduction converts features to a lower dimension. After examining the models using all 68 features, we applied ``Principal component analysis'' (PCA) and ``Univariate linear regression test'' (KBest) for feature reduction and feature selection, respectively. 
PCA  is  a  linear  dimensionality  reduction  technique  that uses  Singular  Value  Decomposition  (SVD)  to  project  the given  data  to  a  lower  dimensional  space. Given  that  we  did  not have  any  estimation  for  the  amount  of  feature  reduction,  we decided  to  consider  the  90\%  of  explained  variance  of  the dataset  as  the  dimension  of  PCA. 
KBest is a  univariate linear   regression   tests for   selecting $k$ best  features using a scoring function. We considerd 25Best features using F statistic as the statistical test between outputs and features for regression. 


\subsection{Hyperparameter Tuning}
Hyperparameter tuning is the process of selecting the best parameters for a model to obtain the optimal results. Grid search is a technique that can be employed to find the optimal parameters of the model through which all combinations of the determined values for parameters are examined. We performed a grid search for hyperparameter tuning on the Random Forest model to find the optimal values. Here is the list of parameters that were tuned for the Random Forest model:
\begin{itemize}
    \item[--] $n\_estimators: [50, 100, 150, 200, 250, 300]$, number of trees in the forests,
    \item[--] $max\_depth:[5, 10, 20, 30, 50]$, maximum number of levels in each decision tree, 
    \item[--] $min\_samples\_split:[2,3,4,5,6,7]$, minimum number of data points placed in a node before the node is split,
    \item[--] $min\_samples\_leaf:[1,2,3,5]$, minimum number of data points allowed in a leaf node, 
    \item[--] $k:range(10,30)$, number of features selected using KBest. 
\end{itemize}

\subsection{Evaluation Metrics}
In order to measure the performance of the regression models, 
%
$RMSE$ and $R^2$ were chosen to evaluate the performance of each regression model.

$R^2$  provides a comparison of total sum of squares of prediction error with total sum of squares of error with mean. The closer the value of $R^2$ to 1 is, the better the regression model will be.
It should be mentioned that $R^2$ is less common used metric for assessing non-linear models~\cite{Spiess2010}.

$RMSE$ can be considered as the standard deviation of the prediction errors. Because it applies a high penalty on large errors, it is beneficial when large errors are unwanted. 

\begin{figure*}[htpb!]
    \begin{subfigure}{0.50\textwidth}
      \includegraphics[width=\textwidth]{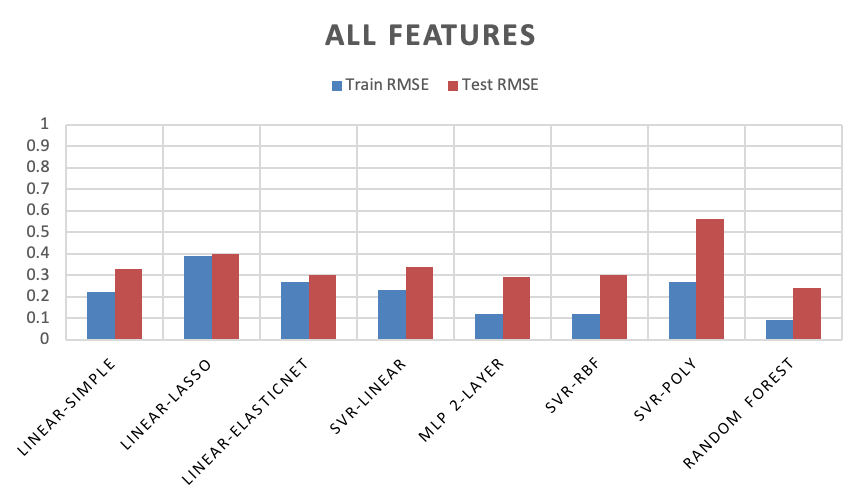}
      \caption{$RMSE$}
      \label{fig:aro:emo-rmse-all}
    \end{subfigure}%
    \begin{subfigure}{0.50\textwidth}
      \includegraphics[width=\textwidth]{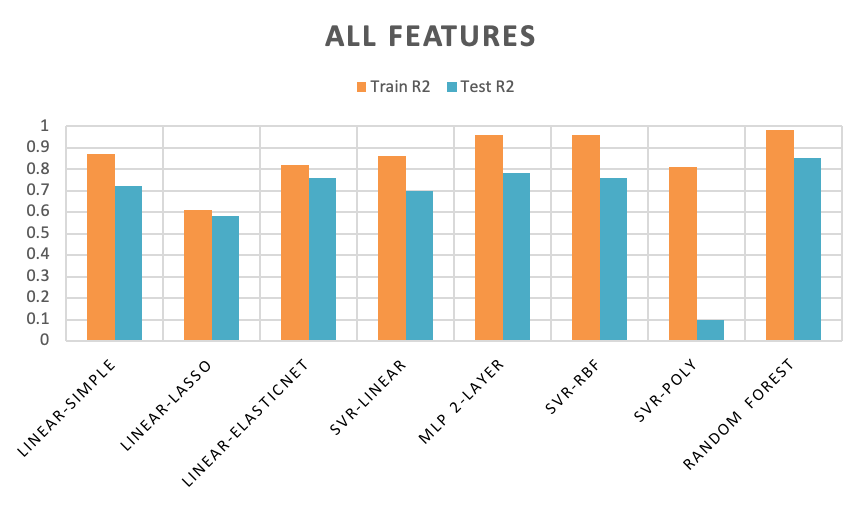}
      \caption{$R^2$}
      \label{fig:aro:emo-r2-all}
    \end{subfigure}%
    
    \begin{subfigure}{0.50\textwidth}
      \includegraphics[width=\textwidth]{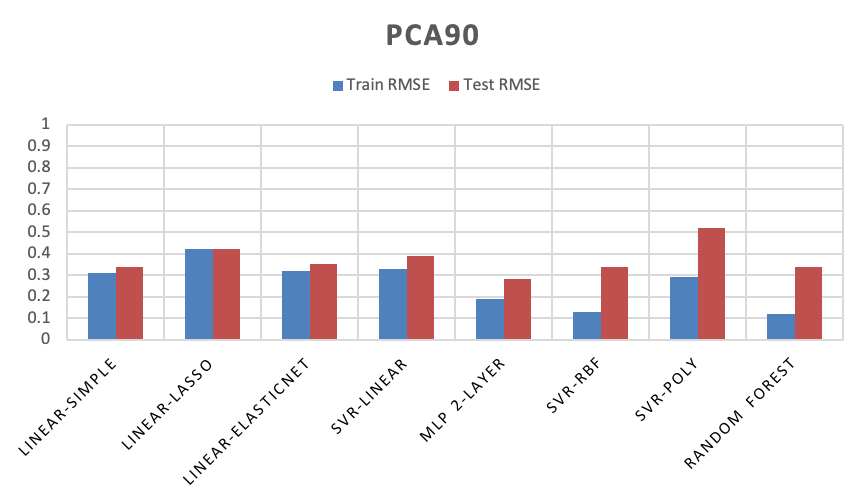}
      \caption{$RMSE$}
      \label{fig:aro:emo-rmse-pca90}
    \end{subfigure}%
    \begin{subfigure}{0.50\textwidth}
      \includegraphics[width=\textwidth]{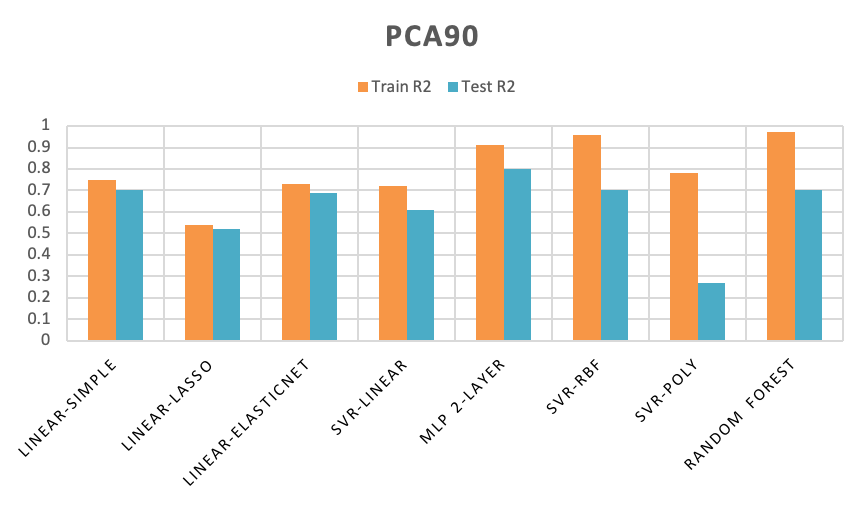}
      \caption{$R^2$}
      \label{fig:aro:emo-r2-pca90}
    \end{subfigure}

        \begin{subfigure}{0.50\textwidth}
      \includegraphics[width=\textwidth]{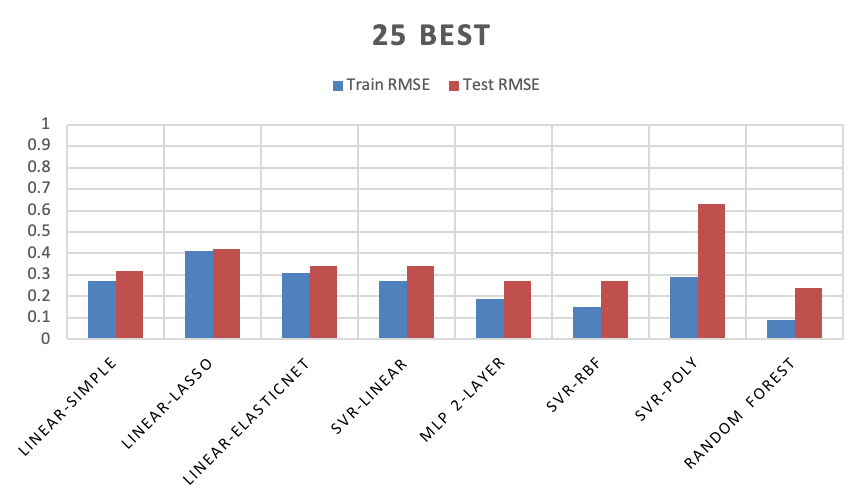}
      \caption{$RMSE$}
      \label{fig:aro:emo-rmse-25best}
    \end{subfigure}%
    \begin{subfigure}{0.50\textwidth}
      \includegraphics[width=\textwidth]{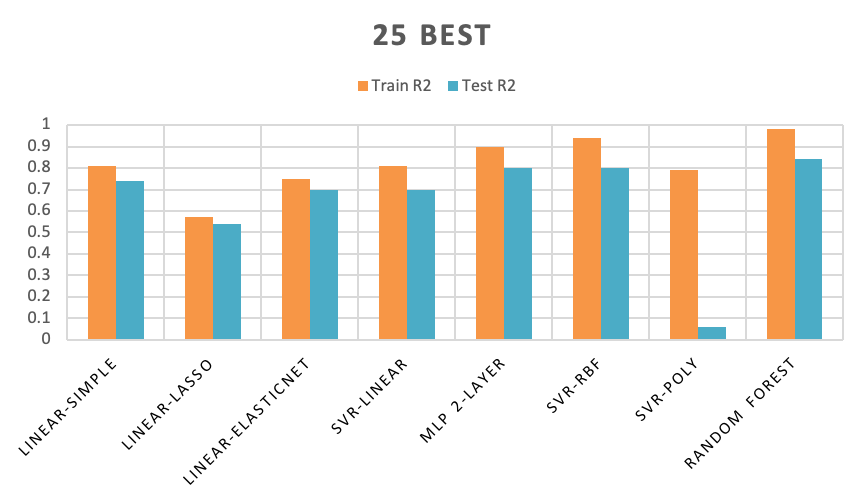}
      \caption{$R^2$}
      \label{fig:aro:emo-r2-25best}
    \end{subfigure}

    \caption{Arousal prediction.}
    \label{fig:aro_allfeatures}
    \vspace{-0.2in}
\end{figure*}

\section{Prediction Results}
\label{sec:results}
This section reports the results of our analysis on building models for predicting perceived emotions. As discussed earlier, emotions perceived or induced from sounds are quantified in terms of two factors: 1) arousal, and 2) valence. We first present the performance of predicting models for arousal, and then discuss our results for predicting valence.

\begin{figure*}[htpb!]
    
    \begin{subfigure}{0.50\textwidth}
      \includegraphics[width=\textwidth]{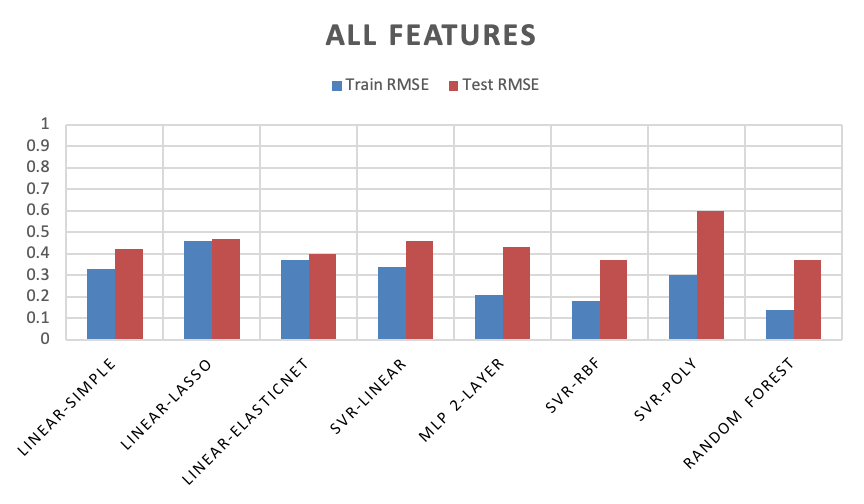}
      \caption{$RMSE$}
      \label{fig:val:emo-rmse-all}
    \end{subfigure}%
    \begin{subfigure}{0.50\textwidth}
      \includegraphics[width=\textwidth]{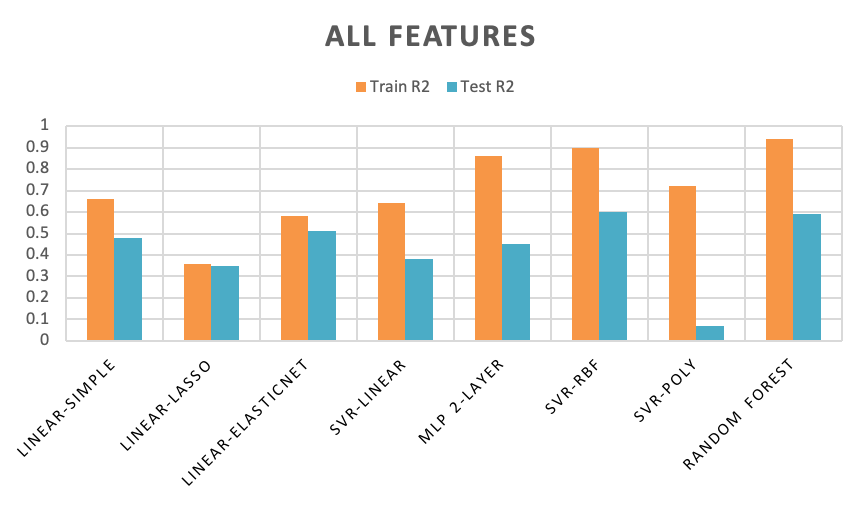}
      \caption{$R^2$}
      \label{fig:val:emo-r2-all}
    \end{subfigure}%
    
    \begin{subfigure}{0.50\textwidth}
      \includegraphics[width=\textwidth]{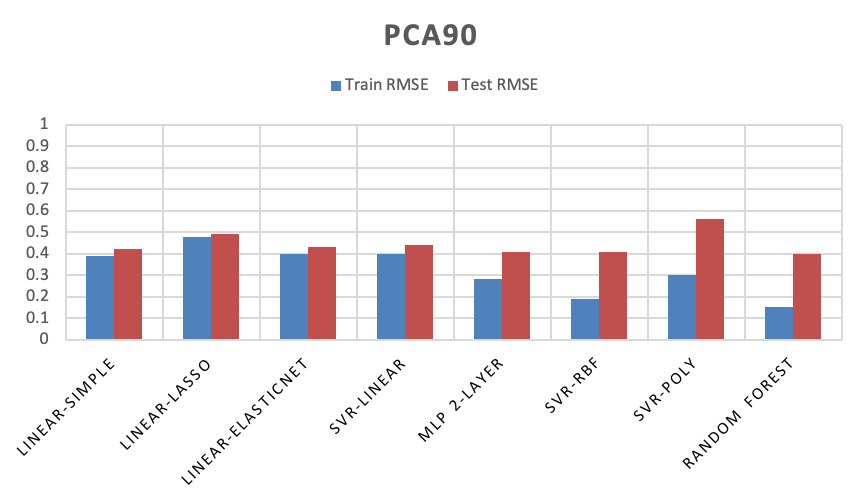}
      \caption{$RMSE$}
      \label{fig:val:emo-rmse-pca90}
    \end{subfigure}%
    \begin{subfigure}{0.50\textwidth}
      \includegraphics[width=\textwidth]{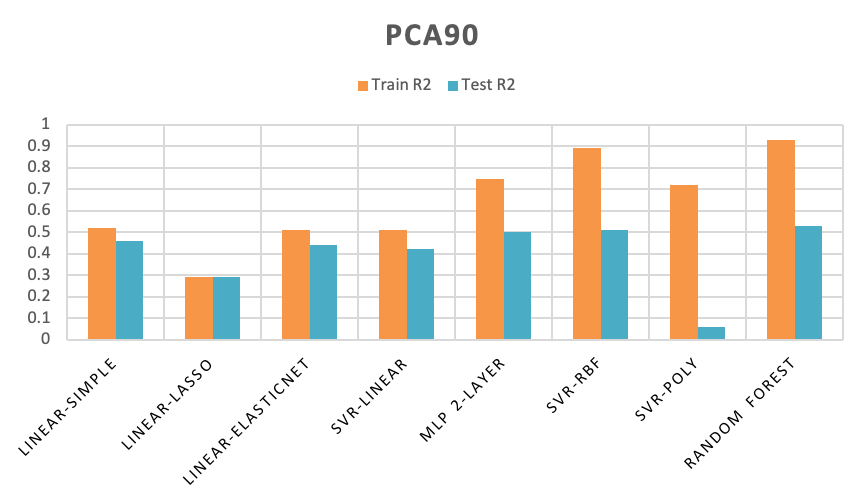}
      \caption{$R^2$}
      \label{fig:val:emo-r2-pca90}
    \end{subfigure}

        \begin{subfigure}{0.50\textwidth}
      \includegraphics[width=\textwidth]{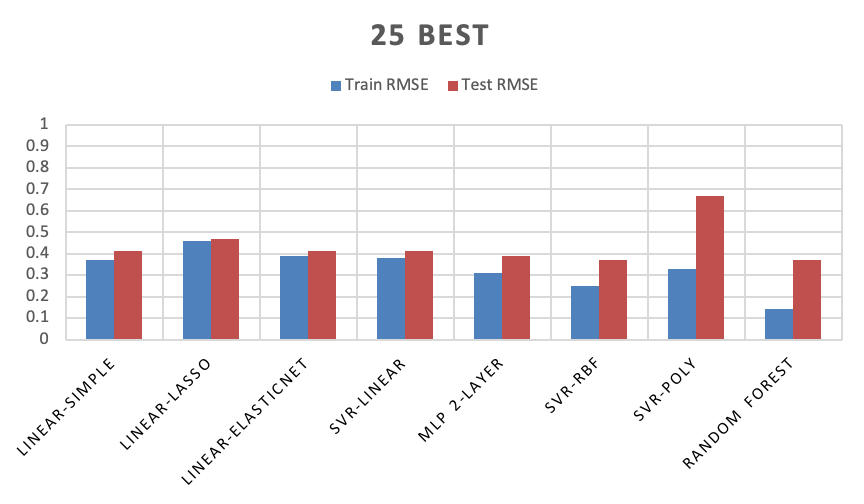}
      \caption{$RMSE$}
      \label{fig:val:emo-rmse-25best}
    \end{subfigure}%
    \begin{subfigure}{0.50\textwidth}
      \includegraphics[width=\textwidth]{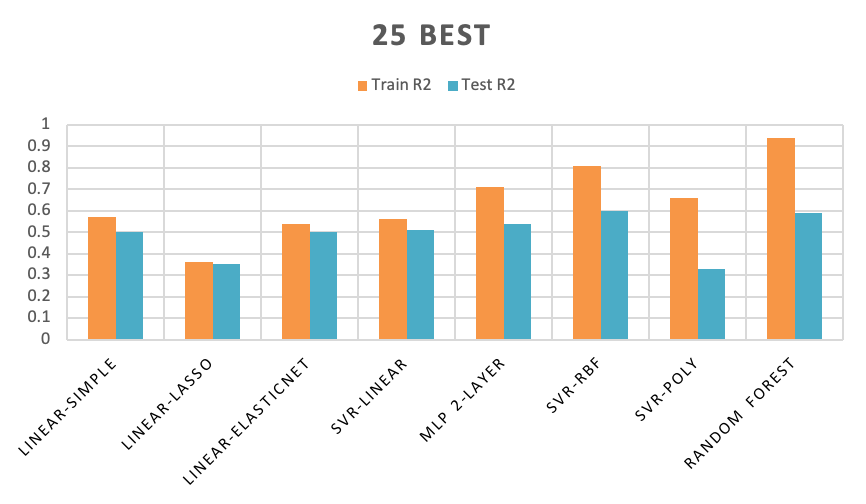}
      \caption{$R^2$}
      \label{fig:val:emo-r2-25best}
    \end{subfigure}

    \caption{Valence prediction.}
    \label{fig:val_allfeatures}
    \vspace{-0.12in}
\end{figure*}

\subsection{Predicting Perceived Arousal of Sounds}

The barplots shown in Figure \ref{fig:aro_allfeatures} summarize the results for predicting arousal. We report the results for all features, PCA 90, and 25 Best.  

\subsubsection{All Features}

Using the MIR toolbox in Matlab, we extracted 68 acoustic features from the Emo-Soundscape dataset. As illustrated in Figure \ref{fig:aro:emo-rmse-all}, the $RMSE$ values for the train dataset (i.e., the blue bars) are around $0.09$ and $0.39$ obtained by Random Forest and Linear-Lasso, respectively. Likewise, the $RMSE$ values for test dataset (i.e., the red bars) are around $0.24$ for Random Forest and $0.56$ for Support Vector Regression (Polynomial) (SVR-Poly), respectively. 

Similarly for $R^2$ values and according to Figure \ref{fig:aro:emo-r2-all}, for the training datasets the highest value is obtained by Random Forest ($0.98$), and the lowest value is offered by Linear-Lasso ($0.61$). For the test datasets, the highest $R^2$ value was achieved by Random Forest ($0.85$), and the lowest is achieved by Support Vector Regression (Polynomial) (SVR-Poly) ($0.1$).

According to our results, the Random Forest-based models outperformed other models with respect to $RMSE$ and $R^{2}$. The Support Vector Regression (Polynomial) (SVR-Poly) clearly suffered from an overfitting problem because the metrics calculated for the test datasets are very poor.

\subsubsection{PCA 90}

The application of PCA with 90\% variations of the dataset yielded 28 features. Figures \ref{fig:aro:emo-rmse-pca90} and \ref{fig:aro:emo-r2-pca90} illustrate the $RMSE$ and $R^2$ values when PCA with 90\% variation is utilized for prediction of arousal. 

According to Figure \ref{fig:aro:emo-rmse-pca90}, the minimum and maximum $RMSE$ values were obtained for the training dataset by Random Forest ($0.12$) and Linear-Lasso ($0.42$), respectively. Furthermore, for the test dataset, the minimum and maximum $RMSE$ values were obtained by Random Forest ($0.34$) and Linear-Lasso ($0.42$). The results indicate that, even with the reduced number of features using PCA 90\%, the ensemble-based Random Forest models outperformed the other machine learning estimators.  

In terms of $R^2$, we obtained similar results for both training and testing datasets when the features extracted by PCA 90\% are utilized for prediction of arousal. For the training dataset, the Random Forest offered $0.97$ (the highest of all $R^2$ calculated); whereas, Linear-Lasso exhibited a very poor result, $R^2 = 0.54$. On the other hand, the maximum $R^2$ for the test dataset is achieved by MLP 2-layer with $0.8$ and the poorest result is obtained by SVR-Poly ($R^2 = 0.27$). 
The Random Forest exhibits a competitive estimate of $R^2 = 0.7$ for the test dataset. 

Overall, the results indicate that reducing the number of features to 28 by PCA 90\%, the Random Forest model still offers the best results in predicting arousal; whereas, other models such as Linear-Lasso and SVR-Poly produce poor estimates of model fit.

\subsubsection{25 Best}

Furthermore, we studied the performance of the prediction models when the 25 best features were utilized for building the models. Figures \ref{fig:aro:emo-rmse-25best} and \ref{fig:aro:emo-r2-25best} depict the prediction results through $RMSE$ and $R^2$ values.

For the training, the best $RMSE$ value is offered by Random Forest ($0.09$); whereas, the worst performance is exhibited by Linear-Lasso ($0.41$). For the test dataset, the best $RMSE$ value is again provided by Random Forest ($0.24$); whereas, the worst $RMSE$ value is calculated by SVR-Poly ($0.63$).

In terms of the $R^2$ values and for the training dataset, the best performer is again Random Forest with $R^2 = 0.98$ and the worse model is Linear-Lasso with $R^2 = 0.57$. Similarly, for the test dataset, the best performing model is Random Forest with $R^2 = 0.84$ and the worst performing model is again SVR-Poly with $R^2 = 0.06$. 

In compliance with all features and PCA 90\%, the results obtained by the 25 best features indicate that Random Forest is able to build a better prediction model for predicting arousal.


\subsubsection{All-Features vs. PCA 90\% vs. 25 Best}

Taking the best evaluation values for $RMSE$ and $R^2$ into consideration when applying all features, PCA 90\%, and 25 Best, we observe that 1) Random Forest is the dominant model among all the regressors studied, and 2) all features and 25 best features exhibit similar results in predicting arousal, followed by PCA 90\%, which provided slightly weaker estimates when predicting accuracy in comparison to all features and 25 best features. Given the cost of extracting features, the results indicate that building a prediction model based on 25 best features is a better choice, yielding more accurate prediction models.

\subsection{Predicting Valence Perceived from Sounds}

As the second dimension of predicting perceived emotions, we built similar models for prediction of valence. Similar to the study we performed in the previous section, we built predictive models based on all features, PCA 90\%, and 25 Best features. The barplots shown in Figures \ref{fig:val_allfeatures} present the estimates of $RMSE$ and $R^2$ for the valence predictions.

\begin{table*}[!ht]
\caption{Exhaustive search results for the best A/V prediction.}
\centering
\begin{tabular}{|l|l|l|l|l|l|l|l|l|}
\hline
\multicolumn{1}{|c|}{\bf output} &
\multicolumn{1}{|c|}{\bf KBest} &
\multicolumn{1}{|c|}{\bf n\_estimators} &
\multicolumn{1}{|c|}{\bf max\_depth} &
\multicolumn{1}{|c|}{\bf min\_samples\_split} &
\multicolumn{1}{|c|}{\bf min\_samples\_leaf} &
  \begin{tabular}[c]{@{}l@{}}{\bf Train}\\ $RMSE$\end{tabular} &
  \begin{tabular}[c]{@{}l@{}}{\bf Test} \\ $RMSE$\end{tabular} &
  \begin{tabular}[c]{@{}l@{}}{\bf Search} \\ {\bf Time}\end{tabular} \\ \hline
Arousal    & 26    & 200   & 20   & 3   & 1   & 0.09   & 0.25   & 5:13:37   \\ \hline
Valence    & 29    & 100   & 30   & 2   & 1   & 0.13   & 0.37   & 5:17:51   \\ \hline
\end{tabular}
\label{tab:Exhaustive Search Results for the Best A/V Prediction}
\vspace{-0.1in}
\end{table*}

\subsubsection{All Features}

Similar to the process employed in building models for predicting arousal, we first focused on all 68 features for predicting valence. Figures \ref{fig:val:emo-rmse-all} and \ref{fig:val:emo-r2-all} illustrate the barplots for $RMSE$ and $R^2$ values yielded by machine learning models. For the test data, the best $RMSE$ value was obtained by Random Forest ($0.14$) and the worst $RMSE$ value was offered by Linear-Lasso ($0.46$). Similarly, for the test data, the best performing model was Random Forest ($0.37$), whereas the worst prediction model was once again Linear-Lasso ($0.47$). 

In terms of utilizing $R^2$ for model assessment, for the train dataset, the best performing model was Random Forest, offering a high value of $R^2 = 0.94$; The worst prediction model was again Linear-Lasso. Similarly, for the test dataset, the best performing  prediction model was Random Forest with $R^2 = 0.59$, whereas, the worst model was captured by Linear-Lasso ($R^2 = 0.35$).

\subsubsection{PCA 90}

Taking into account the 28 acoustic features of sounds for the purpose of predicting valence, we built similar prediction models whose performance are visualized through Figures \ref{fig:val:emo-rmse-pca90} and  \ref{fig:val:emo-r2-pca90}. 

According to Figure \ref{fig:val:emo-rmse-pca90}, Random Forest with $0.15$ and $0.4$ $RMSE$ values outperformed other regression models for both training and testing datasets, respectively. The worst model fitting and predictions were produced once again by Linear-Lasso, with $0.48$ and $0.49$ for $RMSE$ of training and testing datasets, respectively. 

Similarly, we observe better performance pronounced in terms of $R^2$ by Random Forest. The $R^2$ values captured by Random Forest for the training and testing datasets are $0.93$ and $0.53$, respectively; whereas, the worst performing model is introduced again by Linear-Lasso with $R^2$ values of $0.29$ for both training and testing datasets, respectively.

\subsubsection{25 Best}

We obtained similar results when building predictive models using the 25 best features. According to Figures \ref{fig:val:emo-rmse-25best} and \ref{fig:val:emo-r2-25best}, the best model achieving the lowest $RMSE$ values was Random Forest, with $RMSE$ values of $0.14$ and $0.37$ for both training and test datasets, respectively; whereas, the worst performing model is built by Linear-Lasso and SVR-Poly. The $RMSE$ values calculated by Linear-Lasso for training and testing datasets are $0.46$ and $0.47$, respectively. Furthermore, $RMSE$ values captured by SVR-Poly for training and testing datasets are $0.33$ and $0.67$, respectively. 

A similar result is observed for $R^2$. The model built based on Random Forest provided the highest $R^2$ values for training and testing as $0.94$ and $0.59$, respectively. Whereas, the worst performing model is once again introduced by Linear-Lasso and SVR-Poly with $0.36$ and $0.66$ for training datasets, respectively. Moreover, the $R^2$ values for test datasets computed by Linear-Lasso and SVR-Poly are $0.35$ and $0.33$, respectively. 

\subsubsection{All Features vs. PCA 90\% vs. 25 Best}

Looking at the performance of Random Forest-based prediction models depicted in Figure \ref{fig:val_allfeatures}, we notice that the performance of models based on all features and 25 best features are competitive and similar. As a result, we believe it would be a better option to build models based on a reduced set of acoustic features of 25 than building models based on all features. 

In compliance with the results we obtained for predicting arousal, we observe that models based on Random Forest also outperformed other regression algorithms with a better accuracy for predicting valence. Therefore, in the following sections, we compare the performance of Random Forest models built using 25 best features in predicting arousal and valence and investigate how different or difficult it is to predict these two dimensions of emotion.

\subsection{Predicting Arousal vs. Valence}

Arousal and valence are the two primary dimensions of emotions and thus are utilized in building models. In other words, the better the prediction for arousal and valence, the more accurate model is achieved for emotion prediction perceived from sounds. Given the fact that the best models are built using Random Forest on top of 25 best features, we compared the performance of predicting arousal and valence using the prediction data captured by Random Forest using 25 best features. Figures \ref{fig:aro:emo-rmse-25best} and \ref{fig:aro:emo-r2-25best} illustrate the prediction accuracy for arousal using Random Forest and 25 best features; whereas, Figures \ref{fig:val:emo-rmse-25best} and \ref{fig:val:emo-r2-25best} depict the same metrics for valence.

As seen in the figures and evident from model fitting evaluation metrics, we were able to fit a slightly more accurate model for arousal than valence. The numerical values for $RMSE$ values for training and testing as well as for $R^2$ values for training and testing captured for arousal are $0.09$, $0.24$, $0.98$, and $0.84$. On the other hand, the $RMSE$ values for training and testing as well as for $R^2$ values for training and testing captured for valence are $0.14$, $0.37$, $0.94$, and $0.59$.

Overall the $RMSE$ values are smaller for models built for predicting arousal while the $R^2$ values are greater than those computed for valence. The results indicate that fitting a good model and predicting valence is harder than building a good model for predicting arousal. The results make intuitive sense because modeling and predicting pleasures (i.e., valence) is much harder than modeling excitements (i.e., arousal)~\cite{Trochidis2013}. More specifically, it is hard to infer whether a sound expresses a positive or negative affect, but it is much easier to conclude that the sound is exciting or dull.

\begin{table}[htpb!]
      \begin{center}
\caption{The best and common features for arousal/valence.}
\begin{tabular}{|l|l|}
\hline
\multicolumn{1}{|c|}{\bf Arousal} & \multicolumn{1}{|c|}{\bf Valence} \\ \hline
\multicolumn{2}{|l|}{dynamics\_rms\_mean}                 \\ \hline
\multicolumn{2}{|l|}{dynamics\_rms\_std}                  \\ \hline
\multicolumn{2}{|l|}{rhythm\_fluctuationmax\_peakposmean} \\ \hline
\multicolumn{2}{|l|}{rhythm\_pulseclarity\_mean}          \\ \hline
\multicolumn{2}{|l|}{spectral\_centroid\_std}             \\ \hline
\multicolumn{2}{|l|}{spectral\_flatness\_mean}            \\ \hline
\multicolumn{2}{|l|}{spectral\_flatness\_std}             \\ \hline
\multicolumn{2}{|l|}{spectral\_irregularity\_mean}        \\ \hline
\multicolumn{2}{|l|}{spectral\_mfcc\_mean\_2}             \\ \hline
\multicolumn{2}{|l|}{spectral\_mfcc\_mean\_8}             \\ \hline
\multicolumn{2}{|l|}{spectral\_novelty\_mean}             \\ \hline
\multicolumn{2}{|l|}{spectral\_novelty\_std}              \\ \hline
\multicolumn{2}{|l|}{spectral\_rolloff85\_mean}           \\ \hline
\multicolumn{2}{|l|}{spectral\_rolloff85\_std}            \\ \hline
\multicolumn{2}{|l|}{spectral\_rolloff95\_mean}           \\ \hline
\multicolumn{2}{|l|}{spectral\_roughness\_mean}           \\ \hline
\multicolumn{2}{|l|}{spectral\_spread\_mean}              \\ \hline
\multicolumn{2}{|l|}{timbre\_lowenergy\_std}              \\ \hline
\multicolumn{2}{|l|}{timbre\_spectralflux\_mean}          \\ \hline
\multicolumn{2}{|l|}{timbre\_spectralflux\_std}           \\ \hline
spectral\_mfcc\_mean\_4     & spectral\_mfcc\_std\_4      \\ \hline
spectral\_mfcc\_std\_10     & spectral\_mfcc\_std\_6      \\ \hline
spectral\_mfcc\_std\_12     & spectral\_centroid\_mean    \\ \hline
spectral\_mfcc\_std\_13     & spectral\_kurtosis\_mean    \\ \hline
spectral\_spread\_std       & spectral\_kurtosis\_std     \\ \hline
timbre\_lowenergy\_mean     & spectral\_skewness\_mean    \\ \hline
                            & tonal\_hcdf\_mean           \\ \hline
                            & tonal\_mode\_std            \\ \hline
                            & rhythm\_tempo\_std          \\ \hline
\end{tabular}
\label{tab:bestfeatures}
 \end{center}
 \vspace{-0.2in}
\end{table}

\subsection{Exhaustive Search for Hyperparameter Tuning}
According to Figures \ref{fig:aro_allfeatures} and \ref{fig:val_allfeatures}, it is clear that the Random Forest outperforms other prediction models. Therefore, Random Forest was selected as the ultimate model to be tuned. Table \ref{tab:Exhaustive Search Results for the Best A/V Prediction} reports the results of tuning five parameters separately for arousal and valence predictions. The best test $RMSE$ value for arousal prediction is 0.25, which is captured using 26 Best features. These values are close to the results obtained by 25 Best features. Considering the valence prediction, the best performance achieved is 0.37 for $RMSE$ using 29 Best features. The 26 Best features for arousal prediction and 29 Best features for valence are reported in Table \ref{tab:bestfeatures}. It is noticeable that 20 features are common in best features for predicting arousal and valence.

\section{Conclusion and Future Work}
\label{sec:conclusion}


This paper focuses on emotions perceived from sounds and explores whether it is possible to predict perceived emotions accurately through machine learning algorithms. To do so, we extracted acoustic features of a given repository of sounds, called Emo-Soundscape, using the MATLAB MIRtoolbox. We then built regression-based machine learning models in which the acoustic features were utilized to build the models. 

Exploring possible regression machine learning algorithms, we observed that ensemble-based learning and in particular Random Forest offer better predictions for both arousal and valence. More specifically, for predicting arousal and for training and testing datasets, we obtained $RMSE$ values of $0.09$ and $0.25$, respectively; whereas, for valence, the $RMSE$ values were $0.13$ and $0.37$ for training and testing datasets.

This study is the first step towards modeling emotions perceived from sounds and through arousal and valence. Additional studies are needed to investigate whether other types of sound features other than acoustic features, could be also useful for prediction purposes. Furthermore, this paper explored conventional machine learning algorithms. The emerging approaches in deep learning analysis might provide better predictions for perceived emotions.  

\vspace{-0.05in}
\section*{Acknowledgment}
This research work is supported by National Science Foundation (NSF) under Grant No: 1564293.

\bibliographystyle{IEEEtran}
\bibliography{IEEEfull,sample-base}

\end{document}